\documentclass[aps,prl,english,twocolumn,showpacs,preprintnumbers,amsmath,amssymb,floatfix]{revtex4}
\usepackage[T1]{fontenc}
\usepackage[latin9]{inputenc}
\usepackage{graphicx}
\usepackage{amssymb}
\usepackage{babel}

\makeatother

\makeatother

\begin{document}

\title{Gravitation of finite range and the accelerated expansion of the universe.}

\author{Bo E. Sernelius}

\affiliation{Division of Theory and Modeling, Department of Physics, Chemistry
and Biology, Link\"{o}ping University, SE-581 83 Link\"{o}ping, Sweden}

\email{bos@ifm.liu.se}

\begin{abstract}
In an earlier work \cite {GravSer} we showed that the gravitational
interaction could be reproduced as a retarded dispersion interaction
(Casimir interaction) between particles composed of hypothetical particles
having harmonic oscillator interactions.  Here we derive the modification
of the force when the vacuum modes have a small but finite temperature. 
The resulting interaction is of finite range. We give the result on analytical
form.

\end{abstract}

\maketitle
In a previous work \cite {GravSer} we opened up for the possibility that
gravity, considered to be one of the fundamental interactions in nature, is
not a fundamental interaction after all.  In our model, gravity is a
dispersion interaction derived from more fundamental forces between
particles having harmonic oscillator interactions.  The dispersion forces
are obtained from the change in the zero-point energy of the normal modes
of the interaction fields in the system when the position of the composite
particles are changed.  This is in complete analogy with the
electromagnetic dispersion forces between atoms \cite{CasPol,Ser}.  In the
electromagnetic case there are two regions: the van der Waals region for
moderate distances; the Casimir region for larger distances.  We obtain
corresponding regions in our model but with the proper choice of
characteristic frequency of the composite particle the van der Waals region
will never be observed; the range is smaller than the diameter of the
composite particle.  The dispersion force will have the $r^{ - 2}$
dependence for all distances just as the gravitational force has.

The reason we started looking for an alternative origin of gravity is
two-fold.  One reason is that one has not been able to verify the existence
of the graviton, the quantum of the gravitational field.  The second is
that recent experiments \cite{Tur} have pointed to an accelerated expansion
of the universe.  "Vacuum energy is both the most plausible explanation and
the most puzzling possibility", according to Ref. \cite{Tur}.  Another
possible explanation could be found in a modified gravity. In the present 
work we pursue along this second path.

Big Bang occurred 13.7 billion years ago.  Soon after the formation of
atoms, matter and electromagnetic radiation decoupled.  Since then the
cosmic background radiation has expanded, in the same way as the rest of
the universe, with a factor of approximately 11 000.  All wave lengths have
increased with, and the temperature of the cosmic background radiation has
decreased with, this factor.  The present temperature has been found to be
2.7 K. In analogy to this the radiation associated with the interaction
between the fundamental particles in our model decoupled from matter at
around the time when the composite particles were formed.  This happened
much earlier than the time of atom formation.  Consequently the temperature
of this cosmic background radiation is much lower than that of the
electromagnetic counterpart.

Here we will derive the dispersion force at finite temperature and
demonstrate that this force has a finite range.  There is a temperature
correction factor that goes towards zero when the distance goes towards
infinity.  The normal modes involved in the dispersion forces are of two
types.  One type is bound to the particles, the atoms in the
electromagnetic case and the composite particles in the case of our model;
it affects the forces only at close range, in the van der Waals range.  The
other type is a vacuum mode, modified by the boundary conditions set up by
the presence of the atoms or composite particles; this type of mode
dominates at large separations, in the Casimir range; it was the only type
of mode in the classical work \cite{Cas} by Casimir on the attraction
between two perfectly conducting plates.  This second type should be
populated in accordance with the temperature of the cosmic background
radiation.  The first type, which we are not concerned with here, could be
populated differently.

We found \cite {GravSer} the interaction potential at zero temperature is
\begin{equation}
\begin{array}{l}
 V\left( r \right) = - \frac{\hbar }{{2\pi }}\int\limits_0^\infty {d\omega
 \alpha _1 \left( {i\omega } \right)\alpha _2 \left( {i\omega } \right)e^{
 - {{2\omega r} \mathord{\left/ {\vphantom {{2\omega r} c}} \right. 
 \kern-\nulldelimiterspace} c}} \left[ {3 - 6\left( {{{\omega r}
 \mathord{\left/ {\vphantom {{\omega r} c}} \right. 
 \kern-\nulldelimiterspace} c}} \right)} \right.} \\
 \,\,\,\,\,\,\,\,\,\,\,\,\,\,\,\,\,\,\,\,\,\,\,\,\,\,\,\,\,\,\,\,\,\,\left. 
 { + \frac{{17}}{2}\left( {{{\omega r} \mathord{\left/ {\vphantom {{\omega
 r} c}} \right.  \kern-\nulldelimiterspace} c}} \right)^2 -
 \frac{9}{2}\left( {{{\omega r} \mathord{\left/ {\vphantom {{\omega r} c}}
 \right.  \kern-\nulldelimiterspace} c}} \right)^3 + \frac{3}{2}\left(
 {{{\omega r} \mathord{\left/ {\vphantom {{\omega r} c}} \right. 
 \kern-\nulldelimiterspace} c}} \right)^4 } \right] \\
 \,\,\,\,\,\,\,\,\,\,\,\,\, = - \frac{\hbar }{{2\pi }}\alpha _1 \left( 0
 \right)\alpha _2 \left( 0 \right)\int\limits_0^\infty {d\omega e^{ -
 {{2\omega r} \mathord{\left/ {\vphantom {{2\omega r} c}} \right. 
 \kern-\nulldelimiterspace} c}} \left[ {3 - 6\left( {{{\omega r}
 \mathord{\left/ {\vphantom {{\omega r} c}} \right. 
 \kern-\nulldelimiterspace} c}} \right)} \right.} \\
 \,\,\,\,\,\,\,\,\,\,\,\,\,\,\,\,\,\,\,\,\,\,\,\,\,\,\,\,\,\,\,\,\,\,\,\left.
  { + \frac{{17}}{2}\left( {{{\omega r} \mathord{\left/ {\vphantom {{\omega
 r} c}} \right.  \kern-\nulldelimiterspace} c}} \right)^2 -
 \frac{9}{2}\left( {{{\omega r} \mathord{\left/ {\vphantom {{\omega r} c}}
 \right.  \kern-\nulldelimiterspace} c}} \right)^3 + \frac{3}{2}\left(
 {{{\omega r} \mathord{\left/ {\vphantom {{\omega r} c}} \right. 
 \kern-\nulldelimiterspace} c}} \right)^4 } \right], \\ \end{array}
\end{equation}
where we have used the fact that the exponential factor drops off so fast
that only the static polarizabilities enter and can be brought outside the
integral.  Making the replacement $\alpha _i
\left( 0 \right) = m_i \sqrt {{{32\pi \gamma } \mathord{\left/ {\vphantom
{{32\pi \gamma } {25\hbar c}}} \right.  \kern-\nulldelimiterspace} {25\hbar
c}}}; i=1,2$ of the static polarizabilities gives
\begin{equation}
\begin{array}{l}
 V\left( r \right) = - \gamma m_1 m_2 \left( {\frac{4}{5}} \right)^2
 \frac{1}{c}\int\limits_0^\infty {d\omega e^{ - {{2\omega r}
 \mathord{\left/ {\vphantom {{2\omega r} c}} \right. 
 \kern-\nulldelimiterspace} c}} \left[ {3 - 6\left( {{{\omega r}
 \mathord{\left/ {\vphantom {{\omega r} c}} \right. 
 \kern-\nulldelimiterspace} c}} \right)} \right.} \\
 \,\,\,\,\,\,\,\,\,\,\,\,\,\,\,\,\,\,\,\,\,\,\,\,\,\,\,\,\,\,\,\,\,\,\left. 
 { + \frac{{17}}{2}\left( {{{\omega r} \mathord{\left/ {\vphantom {{\omega
 r} c}} \right.  \kern-\nulldelimiterspace} c}} \right)^2 -
 \frac{9}{2}\left( {{{\omega r} \mathord{\left/ {\vphantom {{\omega r} c}}
 \right.  \kern-\nulldelimiterspace} c}} \right)^3 + \frac{3}{2}\left(
 {{{\omega r} \mathord{\left/ {\vphantom {{\omega r} c}} \right. 
 \kern-\nulldelimiterspace} c}} \right)^4 } \right] \\
 \,\,\,\,\,\,\,\,\,\,\,\, = - {{\gamma m_1 m_2 } \mathord{\left/ {\vphantom
 {{\gamma m_1 m_2 } r}} \right.  \kern-\nulldelimiterspace} r},  \\
 \end{array}
\end{equation}
where $\gamma$ is the gravitational constant. The force is obtained as
\begin{equation}
\begin{array}{l}
 F\left( r \right) = - \frac{{\partial V}}{{\partial r}} \\
 \,\,\,\,\,\,\,\,\,\,\,\,\, = - \gamma m_1 m_2 \left( {\frac{4}{5}}
 \right)^2 \frac{1}{{cr}}\int\limits_0^\infty {d\omega e^{ - {{2\omega r}
 \mathord{\left/ {\vphantom {{2\omega r} c}} \right. 
 \kern-\nulldelimiterspace} c}} \left[ { - 12\left( {{{\omega r}
 \mathord{\left/ {\vphantom {{\omega r} c}} \right. 
 \kern-\nulldelimiterspace} c}} \right)} \right.} \\
 \,\,\,\,\,\,\,\,\,\,\,\,\,\,\,\,\,\,\,\,\,\,\,\,\,\,\,\,\,\,\,\,\,\,\,\,\,\,\,\,\,\,\,\,\,\,\,\,\,\,\,\,\,\,\,\,\,
 + 29\left( {{{\omega r} \mathord{\left/ {\vphantom {{\omega r} c}} \right. 
 \kern-\nulldelimiterspace} c}} \right)^2 - \frac{{61}}{2}\left( {{{\omega
 r} \mathord{\left/ {\vphantom {{\omega r} c}} \right. 
 \kern-\nulldelimiterspace} c}} \right)^3 \\
 \,\,\,\,\,\,\,\,\,\,\,\,\,\,\,\,\,\,\left. 
 {\,\,\,\,\,\,\,\,\,\,\,\,\,\,\,\,\,\,\,\,\,\,\,\,\,\,\,\,\,\,\,\,\,\,\,\,\,\,\,\,
 + 15\left( {{{\omega r} \mathord{\left/ {\vphantom {{\omega r} c}} \right. 
 \kern-\nulldelimiterspace} c}} \right)^4 - 3\left( {{{\omega r}
 \mathord{\left/ {\vphantom {{\omega r} c}} \right. 
 \kern-\nulldelimiterspace} c}} \right)^5 } \right] \\
 \,\,\,\,\,\,\,\,\,\,\,\,\, = - {{\gamma m_1 m_2 } \mathord{\left/
 {\vphantom {{\gamma m_1 m_2 } {r^2 }}} \right.  \kern-\nulldelimiterspace}
 {r^2 }}. \\ \end{array}
\end{equation}
At finite temperature the integral turns into a summation \cite{Ser}
\begin{equation}
\int\limits_0^\infty {d\omega } \to \frac{{2\pi }}{{\hbar \beta
}}\sum\limits_{n = 0}^\infty {} ;\quad \omega _n = \frac{{2\pi n}}{{\hbar
\beta }},
\end{equation}
where the $n=0$ term should be multiplied by a factor of one half. In the 
present case we do not have to mind this since this term vanishes. We 
have
\begin{equation}
\begin{array}{l}
 F\left( r \right) = - \gamma m_1 m_2 \left( {\frac{4}{5}} \right)^2
 \frac{{2\pi }}{{\hbar \beta cr}}\sum\limits_{n = 0}^\infty {e^{ -
 {{2\omega _n r} \mathord{\left/ {\vphantom {{2\omega _n r} c}} \right. 
 \kern-\nulldelimiterspace} c}} \left[ { - 12\left( {{{\omega _n r}
 \mathord{\left/ {\vphantom {{\omega _n r} c}} \right. 
 \kern-\nulldelimiterspace} c}} \right)} \right.} \\
 \,\,\,\,\,\,\,\,\,\,\,\,\,\,\,\,\,\,\,\,\,\,\,\,\,\,\,\,\,\,\,\,\,\,\,\,\,\,\,\,\,\,\,\,\,\,\,\,\,\,\,\,\,\,
 + 29\left( {{{\omega _n r} \mathord{\left/ {\vphantom {{\omega _n r} c}}
 \right.  \kern-\nulldelimiterspace} c}} \right)^2 - \frac{{61}}{2}\left(
 {{{\omega _n r} \mathord{\left/ {\vphantom {{\omega _n r} c}} \right. 
 \kern-\nulldelimiterspace} c}} \right)^3 \\
 \,\,\,\,\,\,\,\,\,\,\,\,\,\,\,\,\,\,\left. 
 {\,\,\,\,\,\,\,\,\,\,\,\,\,\,\,\,\,\,\,\,\,\,\,\,\,\,\,\,\,\,\,\,\,\,\,\,\,
 + 15\left( {{{\omega _n r} \mathord{\left/ {\vphantom {{\omega _n r} c}}
 \right.  \kern-\nulldelimiterspace} c}} \right)^4 - 3\left( {{{\omega _n
 r} \mathord{\left/ {\vphantom {{\omega _n r} c}} \right. 
 \kern-\nulldelimiterspace} c}} \right)^5 } \right]. \\ \end{array}
\end{equation}
This summation may be performed analytically and we find the temperature 
correction factor is
\begin{equation}
\begin{array}{l}
 G\left( {r,T} \right) = {{F\left( {r,T} \right)} \mathord{\left/
 {\vphantom {{F\left( {r,T} \right)} {F\left( {r,0} \right)}}} \right. 
 \kern-\nulldelimiterspace} {F\left( {r,0} \right)}} \\
 \,\,\,\,\,\,\,\,\,\,\,\,\,\,\,\,\,\,\, = \left( {\frac{4}{5}} \right)^2
 x\left[ { - 12z^2 } \right.\, + 29z^3 \left( {x + 1} \right) \\
 \,\,\,\,\,\,\,\,\,\,\,\,\,\,\,\,\,\,\,\,\,\,\,\,\,\,\,\,\, -
 \frac{{61}}{2}z^4 \left( {x^2 + 4x + 1} \right) \\
 \,\,\,\,\,\,\,\,\,\,\,\,\,\,\,\,\,\,\,\,\,\,\,\,\,\,\,\,\, + 15z^5 \left(
 {x^3 + 11x^2 + 11x + 1} \right) \\
 \,\,\,\,\,\,\,\,\,\,\,\,\,\,\,\,\,\,\left.  {\,\,\,\,\,\,\,\,\,\,\, - 3z^6
 \left( {x^4 + 26x^3 + 66x^2 + 26x + 1} \right)} \right], \\ \end{array}
\end{equation}
where $y = {{2\pi r} \mathord{\left/ {\vphantom {{2\pi r} {\hbar \beta c}}}
\right.  \kern-\nulldelimiterspace} {\hbar \beta c}}$, $x = \exp \left( { -
2y} \right)$, $z = {y \mathord{\left/
{\vphantom {y {\left( {1 - x} \right)}}} \right. 
\kern-\nulldelimiterspace} {\left( {1 - x} \right)}}$, and $\beta = {1
\mathord{\left/ {\vphantom {1 {k_B T}}}
\right.  \kern-\nulldelimiterspace} {k_B T}}$.  This temperature correction
factor is displayed in Fig.  (1).

\begin{figure}
\includegraphics[width=8cm]{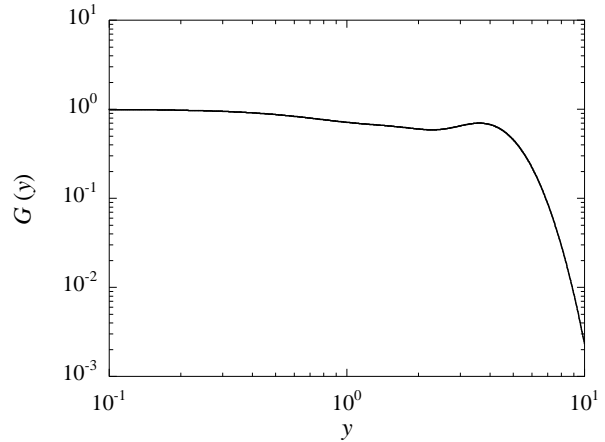}
\caption{The temperature correction factor. See the text for details}
\end{figure}

In summary, we have demonstrated that the gravitational interaction derived 
from particles having harmonic oscillator interaction potentials has a 
finite range if the vacuum modes have a non-zero temperature. We have 
given the temperature correction factor on analytical form. This correction 
factor should have an effect on the expansion of the universe. To be noted 
is that the $r$ and $T$ dependences of the correction factor is in the form 
of the variable $y$. This means that the cut-off distance for gravity 
increases with the same speed as the expansion of the universe.

\begin{acknowledgments}
This research was sponsored by EU within the EC-contract No:012142-NANOCASE
and support from the VR Linn\'{e} Centre LiLi-NFM and from CTS is
gratefully acknowledged.
\end{acknowledgments}

\end{document}